\documentclass[aps,prl,showpacs,amsmath,amssymb,superscriptaddress,footinbib]{revtex4}

\usepackage[english]{babel}
\usepackage{latexsym}
\usepackage{graphics}
\usepackage{subfigure}
\usepackage{epsfig}
\usepackage{amsfonts}
\usepackage{amssymb}
\usepackage{amsmath}
\usepackage{bbm}

\begin{document}

\title{Ultra-cold dipolar gases}

\author{Chiara Menotti} \affiliation{ICFO -- Institut de Ci\`encies
Fot\`oniques, E-08860 Castelldefels, Barcelona, Spain}
\affiliation{CNR-INFM-BEC and Dipartimento di Fisica, Universit\`a di Trento,
I-38050 Povo, Italy}
\author{Maciej Lewenstein} \affiliation{ICFO -- Institut de
Ci\`encies Fot\`oniques, E-08860 Castelldefels, Barcelona, Spain}
\affiliation{ICREA -- Instituci\'o Catalana de Recerca i Estudis Avan\c cats, 
E-08010 Barcelona, Spain} 

\begin{abstract}
We present a concise review of the physics of ultra-cold dipolar
gases, based mainly on the theoretical developments in our own
group. First, we discuss shortly weakly interacting ultra-cold
trapped dipolar gases. Dipolar Bose-Einstein condensates exhibit
non-standard instabilities and the physics of both Bose and Fermi
dipolar gases depends on the trap geometry. We focus then the
second part of the paper on strongly correlated dipolar gases and
discuss ultra-cold dipolar gases in optical lattices. Such gases
exhibit a spectacular richness of quantum phases and metastable
states, which may perhaps be used as quantum memories. We comment
shortly on the possibility of superchemistry aiming at the
creation of dipolar heteronuclear molecules in lattices. Finally,
we turn to ultra-cold dipolar gases in artificial magnetic
fields, and consider rotating dipolar gases, that provide in our
opinion the best option towards the realization of the fractional
quantum Hall effect and quantum Wigner crystals.
 \end{abstract}

\maketitle

 \section{Introduction}

This paper has been presented as an invited lecture at the
International Conference on Recent Progress on Many Body Theories,
RPMBT 2007, held in Barcelona in July 2007, in the session devoted
to ultra-cold atoms. This conference puts traditionally a lot of
emphasis on the development of new methodologies, analytic and
numerical methods for many body problems. The lecture given by M.
Lewenstein was a little different in character: instead of talking
about some new specific method, M. Lewenstein gave a broad review
of the topic, trying to convince the audience  that ultra-cold
dipolar gases provide a fantastic playground to apply modern many
body theory.

This topic belongs to one of the hottest areas of modern AMO
physics, and giving a full review in a one hour lecture is
impossible. M. Lewenstein based his presentation mainly on the
activities of his own group, which fortunately touch most of the
aspects of the physics of ultra-cold dipolar gases (UDG),
mentioning only some selected important contributions from other
groups. In this sense the present paper is not a review in the
strict sense; it is more like a review of important subtopics
within the main topic.

\paragraph{\bf Outline.} 
The outline of this paper is thus the following. It consists of three,
relatively independent parts. In the first part we introduce UDGs, and
argue why they are so interesting and how to realize them in the
laboratory. We sketch very briefly the physics of weakly interacting
trapped dipolar Bose and Fermi gases, and talk about the influence of
the trap geometry on the physical properties of the UDGs.

The second subject of the paper concerns ultra-cold dipolar gases in
optical lattices, that are examples of strongly correlated
systems. Such gases exhibit a spectacular richness of quantum phases
(Mott insulators and insulating checkerboard phase, superfluid and
supersolid phases), as well as an extravagantly large variety of
metastable states, which may perhaps be used as quantum memories. We
comment here shortly on the possibility of superchemistry aiming at
the creation of dipolar heteronuclear molecules in lattices.

In the third and last part,  we turn to ultra-cold dipolar gases
in artificial magnetic  fields and consider rotating dipolar
gases, that provide in our opinion the best option towards the
realization of the fractional quantum Hall effect and quantum
Wigner crystals.

\section{Weakly interacting trapped dipolar gases}

\paragraph{\bf Why dipolar gases?}

Some of the most fascinating experimental and theoretical
challenges of modern atomic and molecular physics arguably concern
ultra-cold dipolar quantum gases \cite{Baranov:2002}. The recent
experimental realization of a quantum degenerate dipolar Bose gas
of Chromium \cite{Griesmaier:2005}, and the progress in trapping
and cooling of dipolar molecules \cite{SpecialIssueEurphysD04}
have opened the path towards ultra-cold quantum gases with
dominant dipole interactions.  In particular just before the
Barcelona RPMBT conference, the group of Tilman Pfau in Stuttgart
realized a UDG of Chromium with dominant  magnetic dipole
interactions, employing a Feshbach resonance to "turn off" the
short range Van der Waals forces \cite{Lahaye}. Several groups
have reported in 2007 enormous progresses in trapping and
manipulating mixtures of different atomic species in an optical
lattice (cf. \cite{sengstock}). Such systems realize the first
step towards the "superchemistry" of UDGs, that we discuss in the
second part of this paper.

\begin{figure}[tbp]
\centering
\includegraphics[width=0.50\textwidth]{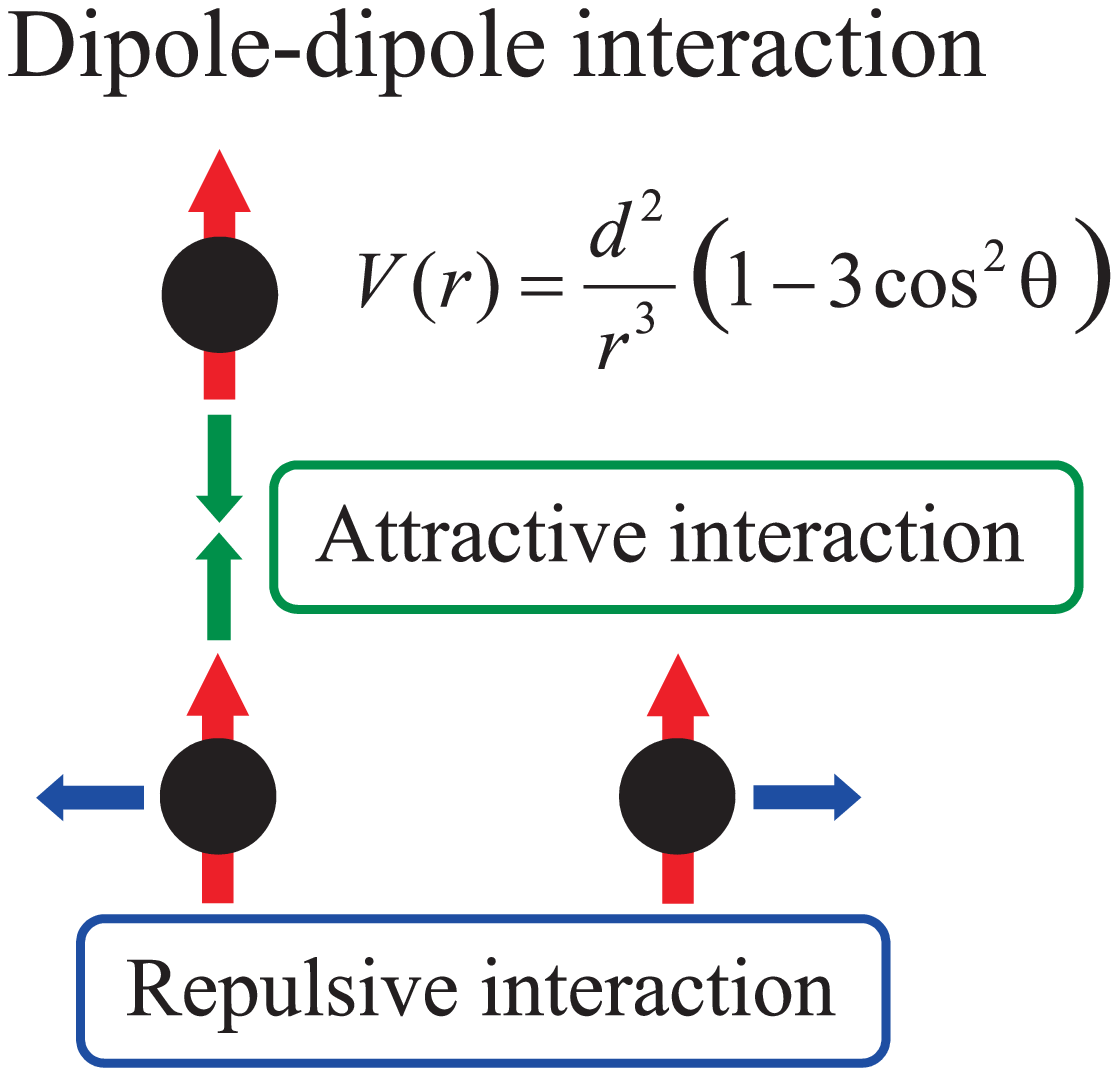}
\caption{Schematic representation of the anisotropic character of
dipole-dipole interaction between dipoles oriented vertically (red
arrows): for relative distances along the orientation of the dipoles
the interaction is attractive (green arrows), while for relative
distances perpendicular to the orientation of the dipoles the
interaction is repulsive (blue arrows).} \label{fig1}
\end{figure}

Why are dipole interactions  interesting? Because of their very
nature. For most of the  systems studied so far, one assumes the
dipole moment to be polarised, i.e. oriented in the same direction
via applying either magnetic or electric fields. In such case the
dipolar potential between two particles is

\begin{equation}
V(r)=\frac{d^2}{r^3}\left(1-3\cos^2\theta\right), \label{dipoles}
\end{equation}
where $r$ is the inter-particle distance, $\theta$ is the angle
between the direction of the dipole moment and the vector
connecting the particles. The interaction is anisotropic and
partially attractive. In particular, if the two dipoles are on top
of one another, they attract themselves, if they are aside, they
repel each other (see Fig.~\ref{fig1}). Locating the dipoles in a
vertical cigar-shaped trap leads to a {\it mainly attractive} gas
that should therefore exhibit collapse. Locating the dipoles in a
horizontal pancake trap leads to a {\it mainly repulsive gas} and
should allow to at least partially stabilise the system.

Several groups have attacked the theory  of the UDGs starting from
1999 (for reviews see \cite{Baranov:2002,Misha2007}). Particularly
important were the pioneering papers by G\'oral, Pfau and Rz\c
a{\.z}ewski \cite{goral}, the L.~You group (cf. \cite{you}),  or the 
G.~Kurizki group \cite{kurizki}.

\paragraph{\bf Trapped dipolar Bose gases}

\begin{figure}[tbp]
\centering
\includegraphics[width=0.50\textwidth]{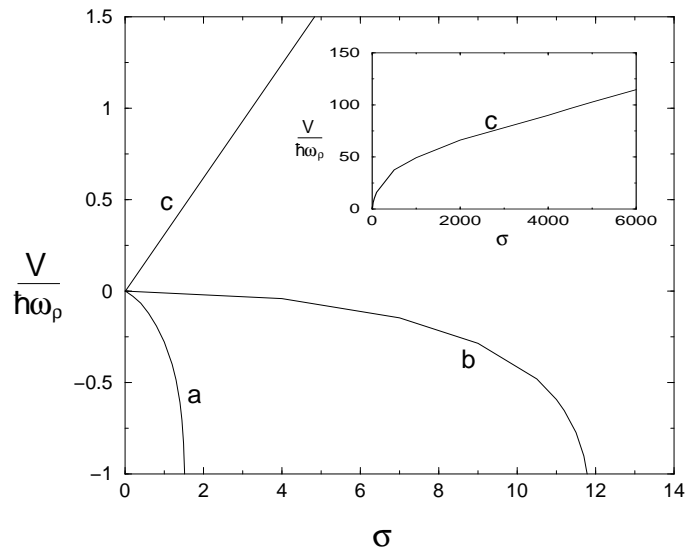}
\caption{Mean dipole-dipole interaction energy $V$ versus $\sigma
\propto Nd^2$ for various regimes: (a) $\lambda=10$, (b) $\lambda=1$
show collapse for a cigar-shaped and isotropic trap, and (c)
$\lambda=0.1$ shows a stable BEC for hard pancake trap. In the inset
the figure (c) is depicted in a larger scale.} \label{fig2}
\end{figure}

The dependence of the physics of trapped dipolar gases on geometry
is quite strong and leads to dramatic effects \cite{santos2000},
as described in the following and summarised in Fig.\ref{fig2}.
Let us for simplicity consider a dipolar gas of $N$ Bose particles
trapped in a harmonic trap, with all of the dipoles $d$ oriented
in the same direction, and assume that the particles interact
dominantly via dipole-dipole interaction. This means, for
instance, that we are dealing with heteronuclear molecules with
sufficiently large electric dipoles, or a Chromium gas with small
magnetic moments interactions, but even smaller $s$-wave
scattering length. At sufficiently low temperatures such a gas
will undergo condensation, and its behaviour is well described by
the Gross-Pitaevskii equation (this fact is by no means obvious).

Let us first consider cigar-shaped ($\omega_\rho\ge \omega_z$) and
"soft" pancake-shaped traps with ($\omega_\rho\le \omega_z$).  There
appears a kind of quantum phase transition as a function of the trap
aspect ratio $\lambda^*=(\omega_\rho/\omega_z)^{1/2}\simeq 0.4$, above
which the sign of the energy of the mean dipole interaction $V$
changes from positive to increasingly negative, as we increase $Nd^2$
for $\lambda^*<\lambda\le 1$, and remains always increasingly negative
for $\lambda>1$.  The condensate becomes more and more cigar-shaped,
until it undergoes a collapse, somewhat similar to what is occurring
for a gas with negative scattering length, i.e.  effectively
attractive Van der Waals interactions \cite{hulet}.

\begin{figure}[tbp]
\centering
\includegraphics[width=0.50\textwidth]{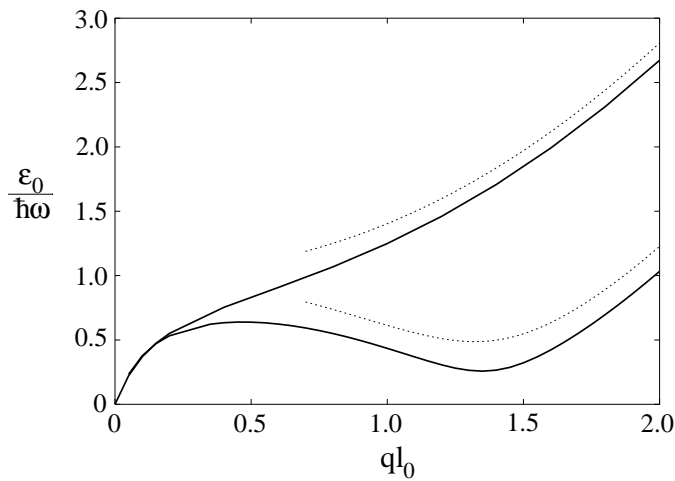} 
\caption{ Dispersion law $\epsilon_0(q)$ for
various values of the ratio between on-site and dipole interaction
strength $\beta$ and $\mu/\hbar\omega$: $\beta=0.53$,
$\mu/\hbar\omega=46$ (upper curve) and $\beta=0.47$,
$\mu/\hbar\omega=54$ (lower curve). The solid curves show the
numerical results, and the dotted curves the result of an analytic
perturbative approach (see \cite{santos2003}).} \label{fig3}
\end{figure}

For hard pancake traps with $\lambda<\lambda^*$, $V$ grows as we
increase $Nd^2$ and the gas is dominantly repulsive. There is no
standard collapse and BEC with much larger values of $N$ are
stable. The condensate aspect ratio decreases with $Nd^2$. For
$\omega_\rho \ll \omega_z$, one can distinguish two regimes:

i) for $\omega_\rho \ll V \ll  \omega_z$, we deal with a quasi-2D
Bose gas with repulsive interactions, which attains radially a
parabolic Thomas-Fermi profile;

ii) for $V\ge \hbar\omega_z$, we deal with the 3D gas in the
Thomas-Fermi regime. Here the gas does feel the attractive part of
the dipolar interactions and undergoes a short wavelength
instability, which leads to a roton-maxon minimum and then
instability in the excitations spectrum \cite{santos2003}
(see Fig.\ref{fig3}).

\paragraph{\bf Trapped dipolar Fermi  gases}

The dependence of the physics of trapped dipolar gases on the
geometry leads also to a quantum phase transition in the case of
Fermi gases \cite{dobrek}. Let us consider again for simplicity a
dipolar gas of $N$ Fermi particles trapped in a harmonic trap,
with all of the dipoles $d$ oriented in the same direction, and
interactions being dominantly of dipole-dipole kind. The question
is whether at sufficiently low temperatures such a gas will
undergo a transition to a superfluid state
(Bardeen-Cooper-Schriefer (BCS) state),  and whether its behaviour
is well described by the BCS equations (again, this latter fact is
by no means obvious).  Pioneering papers on this subject were
written by the groups of L. You and H. Stoof, who have looked at
the possibility of $p$-wave pairing \cite{p-wave}, and the group
of K. Rz\c a\.zewski, who studied the Thomas-Fermi theory
\cite{goral2}. The BCS theory in homogeneous dipolar gas was
investigated in detail by Baranov and Shlyapnikov
\cite{baranov2002}.

\begin{figure}[tbp]
\centering
\includegraphics[width=0.50\textwidth]{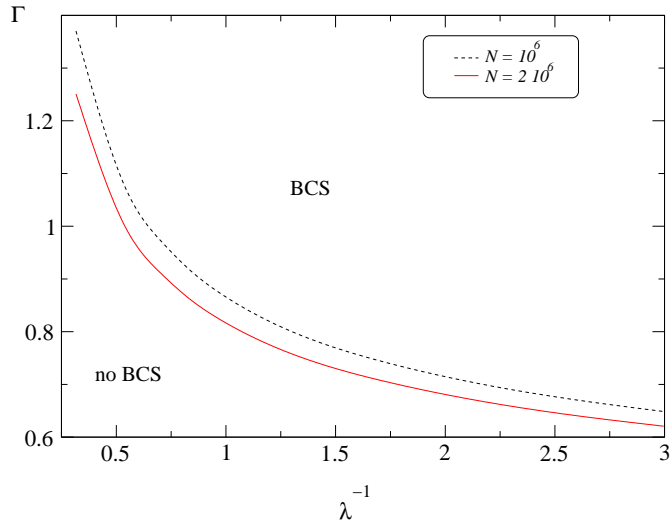}
\caption{ Phase diagram for a trapped dipolar Fermi gas as a function
of $\lambda^{-1}$ showing the critical dipole interactions in units of
Fermi energy $\Gamma$ above which BCS takes place. The upper (lower)
curve corresponds to $N=10^6$ ($N=2 \times 10^6$).} \label{fig4}
\end{figure}

In \cite{dobrek}, we have looked at the BCS transition in a trap,
and have shown indeed the existence of a critical aspect ratio,
similarly as in the case of a Bose gas. The phase diagram is
presented in Fig.~\ref{fig4} as a function of $\lambda^{-1}$ and
dipole interactions in units of Fermi energy $\Gamma$. It can be
viewed in two ways: for a given $\lambda$ the systems undergoes a
transition from the normal to the superfluid state as the dipole
interactions grow. Conversely, for a fixed dipole interactions the
system undergoes the normal-superfluid transition as $\lambda$
decreases. For very small dipole interactions this transitions
occurs in a region of parameters that goes beyond the
applicability of our theory.

\section{Dipolar Bose gases in  optical lattices}

\paragraph{\bf Ultra-cold gases in optical lattices}

Ultra-cold atomic gases in optical lattices (OL) are nowadays the
subject of very intensive studies, since they provide an
unprecedented and unique possibility to study numerous  challenges
of quantum many body physics (for reviews see \cite{bloch,ml}). In
particular, such systems allow to realize various versions of
Hubbard models \cite{zoller}, a paradigmatic example of which is
the Bose-Hubbard model \cite{fisher}. This model exhibits a
superfluid (SF) - Mott insulator (MI) quantum phase transition
\cite{sachdev}, and its atomic realization has been proposed in
the seminal paper of Ref.~\cite{jaksch}, followed by the seminal
experiments of Ref.~\cite{blochex}. Several aspects and
modifications of the SF-MI quantum phase transition, or better to
say crossover \cite{batroumi}, have been  intensively studied
recently  (cf. \cite{ml,Bloch04}).

\paragraph{\bf Ultra-cold dipolar gas in a lattice}

We have proposed to look at the UDG in a 2D lattice in 2002
\cite{goral2002}. The Hamiltonian of the system differs from  the
standard Hubbard-Bose model described by the Hamiltonian

\begin{equation}\label{effham3}
H= -t\sum_i [b_i^\dagger
b_{i-1}+\mathrm{h.c.}]+\frac{U}{2}\sum_in_i(n_i-1)-\mu N,
\end{equation}
where $N=\sum_in_i=\sum_ib_i^\dagger b_i$ is the atom number
operator, $t$ is the hopping term, and $U$ denotes the strength of
on-site interactions. A UDG in a lattice is described by an
extended Bose-Hubbard Hamiltonian

\begin{equation}\label{effham4}
H= -t\sum_i [b_i^\dagger
b_{i-1}+\mathrm{h.c.}]+\frac{U}{2}\sum_in_i(n_i-1) +
\frac{U_1}{2}\sum_{<i,j>}n_in_j +
\frac{U_2}{2}\sum_{<<i,j>>}n_in_j + ... -\mu N,
\end{equation}
where the sum over $\langle i,j \rangle$ pertains to nearest
neighbours, the one over $\langle \langle i,j \rangle\rangle$ to
next-nearest neighbours, etc., and $U_i$ are determined by
dipole-dipole interactions. To a good approximation, assuming that all
dipoles are perpendicular to the plane of the lattice, $U_n=
d^2/r_{ij}^3$, where $r_{ij}$ are the distances between the involved
sites; generally it is given by the expression in Eq.~(\ref{dipoles}).
The resulting model exhibits a rich variety of quantum phases: apart
from the standard superfluid and Mott insulator states, it can form a
checkerboard phase at half filling, or close to it in the parameter
space a supersolid state, i.e. superfluid with a periodic density
modulation in the density and in the order parameter. It can also form
a collapsing state if the interactions are too attractive
\cite{goral2002}.
The possibility of realizing a supersolid state with ultra-cold
atoms is at present particularly attractive because, to our
knowledge, its existence, claimed in $^4$He experiments
\cite{4he}, is still controversial \cite{balibar}. Experiments
with ultra-cold dipolar atoms might provide a much cleaner
environment for the creation and observation of such phases.

\paragraph{\bf Metastable states.} 
Most recently, we have pointed out for the first time that a lattice
system with long-range interactions presents many insulating
metastable states in the low tunneling part of the phase diagram
\cite{menotti}.
The metastable states arise as local minima of the energy. We
access them using a mean-field approach and a time dependent
Gutzwiller Ansatz, which allows to study the dynamics of the
system both in real and imaginary time.

\begin{figure}[tbp]
\centering
\includegraphics[width=0.4\textwidth]{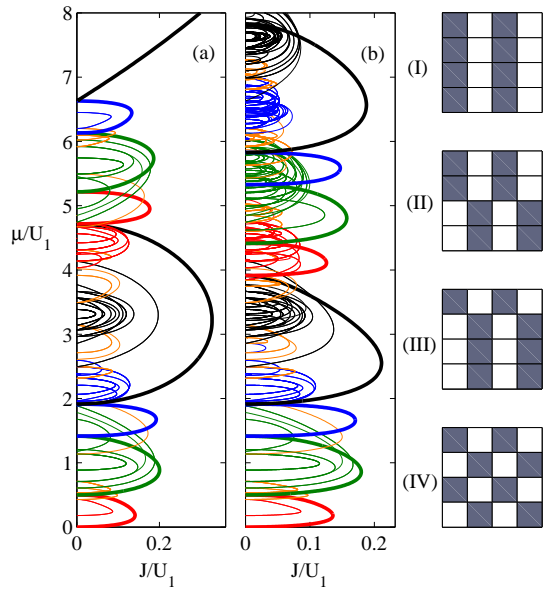}
\caption{ Phase diagram for weak and strong dipole-dipole interaction
and interaction range up to the 4th nearest neighbour: $U/U_{1}=20$
(a) and $U/U_{1}=2$ (b). The thick lines are the ground state lobes,
found (for increasing chemicals potential) for filling factors equal
to all multiples of 1/8. The thin lines are the metastable states,
found at all filling factors equal to multiples of $1/16$. Some of the
metastable configurations at filling factor 1/2 (I to III) and
corresponding ground state (IV).  Empty sites are light and sites
occupied with 1 atom are dark.}
\label{fig5}
\end{figure}

The imaginary time evolution, which mimics dissipation in the
system, converges unambiguously to the ground state of the system
for the Bose-Hubbard model in presence of on-site interaction only.
In the presence of long-range interaction it shows a strikingly
different behaviour and converges often to different
configurations, depending on the exact initial conditions. In this
way, we clearly get a feeling of the existence of metastable
states in the system. In the real time evolution, their stability
is confirmed by typical small oscillations around a local minimum
of the energy.  For all the insulating metastable configurations,
we calculate the insulating lobes in the $J-\mu$ phase space,
using a mean-field perturbative approach. This results in a much
more complex phase diagram, as shown in Fig.~\ref{fig5}.

The metastable states have a finite lifetime due to the tunneling
to different metastable states. We have used a path integral
approach in imaginary time, combined with a dynamical variational
method to estimate this lifetime, which results to be very long
for small tunneling parameter $J$ and large systems.

However, for large systems the number of the metastable states and
the variety of their patterns is so large, and their energy
separation so small, that it turns out to be very difficult to
control the presence of defects. We have checked that by using
superlattices one can prepare the atoms in configurations of
preferential symmetry with very small uncertainty. If a given
configuration corresponds to a metastable state, it survives also
once the superlattice is removed, due to dipole-dipole
interaction.

\begin{figure}[tbp]
\centering
\includegraphics[width=0.65\textwidth]{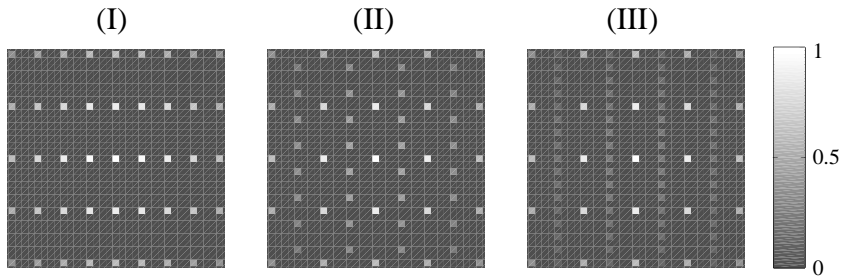}
\caption{Spatial noise correlation patterns for configurations (I) to
(III) in Fig.~\ref{fig1}, assuming a localised gaussian density
distribution at each lattice site.}
\label{fig6}
\end{figure}

For the detection scheme we have presently in mind, it is also
essential (not in line of principle, but practically for present
experimental possibilities) to create a given configuration in a
reproducible way. In fact, the spatially modulated structures
characterising the metastable states can be detected via the
measurement of the noise correlations of the expansion pictures
\cite{altman,bloch_nc,scarola}, which equal the modulus square of
the Fourier transform of the density distribution in the lattice
and is in principle able to recognise the periodic modulations or
the  defects in the density distribution. However, since the signal to
noise ratio required for single defect recognition is beyond the
present experimental possibilities, one should average over a
finite number of different experimental runs producing the same
spatial distribution of atoms in the lattice, and hence accurate
reproducibility is required.
In Fig.~\ref{fig6}, we show the noise correlations for the
metastable configurations at filling factor $1/2$ shown in
Fig.~\ref{fig5}, (I) to (III).

Presently we are studying the possibility of transferring in a
controlled way those systems from one configuration to another.
This, together with the capability of initialising and reading out
the state of the lattice, may make those systems useful for
applications as quantum memories.

\begin{figure}[tbp]
\centering
\includegraphics[width=0.50\textwidth]{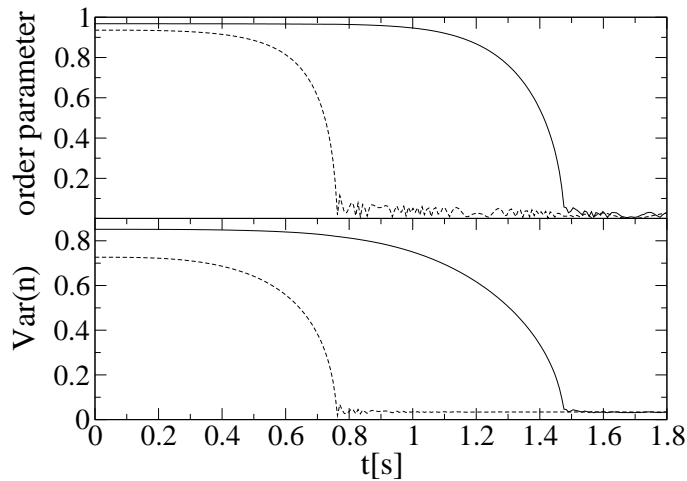}
\caption{Creating a two species Mott insulator in
a real time starting from a superfluid phase of $^{41}{\rm K}$
(solid line) and $^{87}{\rm Rb}$ (dashed line); the upper plot
shows the value of the order parameters $|\langle a_i\rangle|$,
$|\langle b_i\rangle|$ (constant for all lattice sites) for both
species, while the lower one depicts the variance ${\rm
Var}(n)=\sqrt{\langle n^2\rangle - \langle n \rangle^2}$ of the
on-site occupation.} \label{fig7}
\end{figure}

\paragraph{\bf "Superchemistry" of dipolar heteronuclear molecules
in an optical lattice.}
The idea is to create heteronuclear molecules starting from a Mott
insulating phase with exactly one atom per species per site
\cite{hetero}. Polar dimers would be then formed by
photo-association or by using a Feshbach resonance.  With an
appropriate choice of of scattering lengths, such that the two
species do not remain immiscible, one can demonstrate that the
two-species Mott state is the ground state of the system and can
be reached starting by a two-component superfluid and slowly
ramping up the optical lattice potential (see Fig.~\ref{fig7}).

\begin{figure}[tbp]
\centering
\vspace*{0.5cm}
\includegraphics[width=0.50\textwidth]{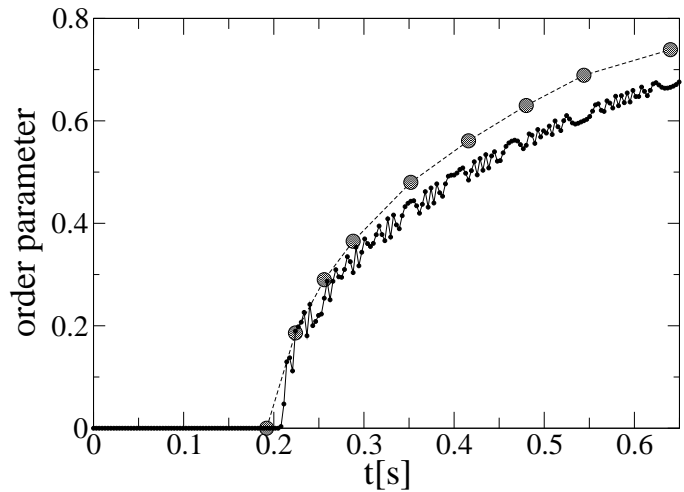}
\caption{Quantum melting of $^{41}{\rm
K}-^{87}{\rm Rb}$ dimers initially in the MI phase towards the
ultra-cold dipolar molecular BEC; the plot shows the time evolution
of the molecular superfluid order parameter $|\langle b_i\rangle|$
(solid line), which is the same for all lattice sites. The dashed
line refers to static calculations of the ground state of the
dipolar molecules placed in the lattice.} \label{fig8}
\end{figure}

Before the creation of molecules, the system is described by a
two-species Bose-Hubbard Hamiltonian with local interactions

\begin{eqnarray}
H=\sum_{\langle i,j\rangle} \left[J_a a_i^{\dag} a_j + J_b
b_i^{\dag} b_j \right] + U_{ab} \sum_i n_{ai}n_{bi} + \frac{1}{2}
\sum_i \left[U_{0a}n_{ai}(n_{ai}-1) +
U_{0b}n_{bi}(n_{bi}-1)\right],
\end{eqnarray}
with $a$ and $b$ denoting the two species, while, after the
creation of the molecules, assuming molecules with non negligible
dipole moment, the Hamiltonian is the one of a single molecular
component gas with long-range interactions, as written in
Eq.~(\ref{effham4}). Finally, this Mott insulating state of
dipolar molecules can be melted to a superfluid heteronuclear
molecular condensate, as shown in Fig.~\ref{fig8}.

\section{\bf Ultra-cold gases in artificial gauge fields}

\paragraph{\bf Rotating ultra-cold gases}
It is well known that rapidly rotating harmonically trapped gases of
neutral atoms exhibit effects completely analogous to charged
particles in uniform magnetic fields (for a recent overview, see for
instance \cite{nuri}). In particular, one should be able to realize
analogues of the fractional quantum Hall effect (FQHE) in such systems
\cite{wilkin,belen}. Particularly interesting in this context are {\em
rotating dipolar gases} (RDG). Bose-Einstein condensates of RDGs
exhibit novel forms of vortex lattices, e.~g., square, stripe- and
bubble-``crystal'' lattices \cite{Cooper:2005}. The stability of these
phases in the lowest Landau level was recently investigated
\cite{Komineas:2006}. We have demonstrated that the quasi-hole gap
survives the large $N$ limit for fermionic RDGs
\cite{Baranov:2005a}. This property makes them perfect candidates to
approach the strongly correlated regime, and to realize Laughlin
liquids (cf. \cite{Girvin}) at filling $\nu=1/3$, and quantum Wigner
crystals at $\nu\le 1/7$ \cite{Fehrmann:2006aa} for a mesoscopic
number of atoms $N\simeq 50-100$. Lately, Rezayi {\it et
al}.~\cite{Rezayi:2005} have shown that the presence of a small amount
of dipole-dipole interactions stabilises the so-called bosonic
Rezayi-Read state at $\nu=3/2$ whose excitations are both fractional
and non-Abelian.

\paragraph{\bf Ordered structures  in rotating Bose gases}
In the recent two years our group has concentrated on studies of small
 samples of rotating atoms using exact diagonalization methods. These
 early studies dealt with the description of ordered structures in
 rotating ultra-cold Bose gases (by looking at the single particle
 density matrix and pair correlation function \cite{nuri}), and by
 studying symmetry breaking in small rotating clouds of trapped
 ultra-cold Bose atoms \cite{nuri2}.  The characterisation of small
 samples of cold bosonic atoms in rotating micro-traps has recently
 attracted increasing interest due to the possibility to deal with a
 few number of particles per site in optical lattices.  In the
 Ref. \cite{nuri} we considered two-dimensional systems of few cold
 Bose atoms confined in a harmonic trap in the $XY$ plane, and
 submitted to strong rotation around the $Z$ axis. By means of exact
 diagonalization, we analysed the evolution of the ground state
 structures as the rotational frequency $\Omega$ increases. Various
 kinds of ordered structures were observed. In some cases, hidden
 interference patterns exhibit themselves only in the pair correlation
 function; in some other cases explicit broken-symmetry structures
 appear that modulate the density.  The standard scenario, valid for
 large systems (i.e., nucleation of vortices into an Abrikosov lattice,
 melting of the lattice, and subsequent appearance of fractional
 quantum Hall type states up to the Lauhglin state), is absent for
 small systems of $N<10$ atoms, and only gradually recovered as $N$
 increases. On the one hand, the Laughlin state in the strong
 rotational regime contains ordered structures much more similar to a
 Wigner crystal or a molecule than to a fermionic quantum liquid. This
 result has some similarities to electronic systems, extensively
 analysed previously. On the other hand, in the weak rotational
 regime, the possibility to obtain equilibrium states whose density
 reveals an array of vortices is restricted to some critical values of
 the rotation frequency $\Omega$.

\paragraph{\bf Rotational symmetry breaking}
In Ref. \cite{nuri2} we have studied the signatures of rotational and
phase symmetry breaking in small rotating clouds of trapped ultra-cold
Bose atoms by looking at the rigorously defined condensate wave
function. Rotational symmetry breaking occurs in narrow frequency
windows, where energy degeneracy between the lowest energy states of
different total angular momentum takes place, and leads to a complex
condensate wave function that exhibits vortices clearly seen as holes
in the density, as well as characteristic vorticities.  Phase symmetry
(or gauge symmetry) breaking, on the other hand, is clearly manifested
in the interference of two independent rotating clouds.

\paragraph{\bf Ultra-cold rotating dipolar Fermi gases}

Armed by the experience on rotating Bose gases with short range
 interactions, we considered in the recent Letter \cite{klaus} a
 system of $N$ dipolar fermions rotating in an axially symmetric
 harmonic trapping potential strongly confined in the direction of the
 axis of rotation. Along this $z$-axis, the dipole moments, as well as
 spins are assumed to be aligned. Various ways of experimental
 realization of ultra-cold dipolar gases are discussed in
 \cite{Baranov:2002}. In case of low temperature $T$ and weak chemical
 potential $\mu$ with respect to the axial confinement $\omega_z$, the
 gas is effectively 2D, and the Hamiltonian of the system in the
 rotating reference frame reads

\begin{equation}{\cal{H}}=\!
     \sum_{j=1}^{N}\!\frac{1}{2M}
     \left(
     \vec{p}_j\!-\!M\Omega\vec{\mathbf{e}}_z\!\times\vec{r}_j
     \right)^2\!
     +\!\frac{M}{2}\!\left({\omega}_0^2-{\Omega}^2\right)r_j^2
     +V_d.
     \label{Hamiltonian}
     \end{equation}
Here, $\omega_{0}\ll \omega_{z}$ is the radial trap frequency,
$\Omega$ is the frequency of rotation, $M$ is the mass of the
particles, $V_d=\sum_{j<k}^N\frac{d^2}{|\vec{r}_j-\vec{r}_k|^3}$ is
the dipolar interaction potential (rotationally invariant with respect
to the $z$-axis), $d$ is the dipole moment, and $\mathbf{r}_{j}=x_{j}
\mathbf{e}_{x}+y_{j}\mathbf{e}_{y}$ is the position vector of the
$j$-th particle.  The first term of \eqref{Hamiltonian} is formally
equivalent to the {Landau} Hamiltonian of particles with mass $M$ and
charge $e$ moving in a constant magnetic field of strength $B=2M\Omega
c/e$ perpendicular to their plane of motion.  The eigenvectors of
${\mathcal{H}}_{\mathrm{Landau}}$ span {Landau} levels (LL) with
energies $\varepsilon_{n}=\hbar\omega_c(n+1/2)$ where
$\omega_c=2\Omega$. We denote by $N_{\mathrm{LL}}=1/2\pi l^{2}$ the
number of states per unit area in each LL, where $l=\sqrt{\hbar
/M\omega_c}$ is the magnetic length. Given a fermionic density $n_f$,
the filling factor $\nu =2\pi l^{2}n_f$ refers to the fraction of
occupied LLs.  Even though the above definition applies to infinite
homogeneous systems, it may be used for finite systems as a suitable
truncation of the Hilbert space at specific angular momenta.  The
second term in \eqref{Hamiltonian} accounts for a rotationally induced
effective reduction of the trap strength. For $\Omega \rightarrow
\omega_{0}$, the confining potential vanishes.

\begin{figure}[h]
\centering
\includegraphics[width=0.49\textwidth,clip]{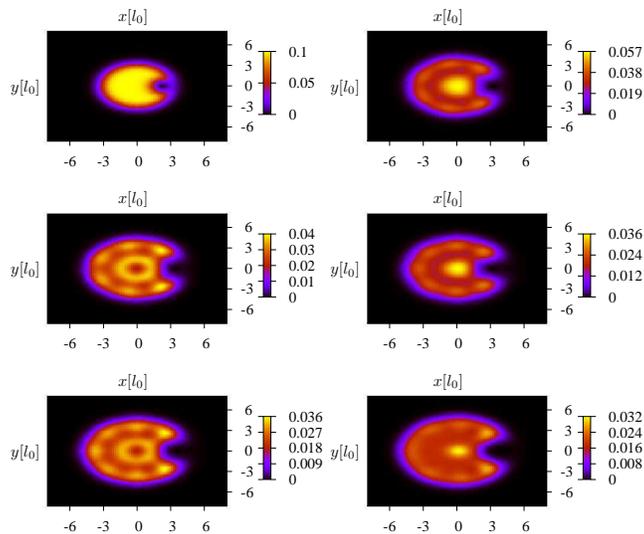}
\caption{Ground state density-density correlation functions
$\hat{\rho}(\vec{r},\,\vec{r}_0)$ for $N=10$ dipolar fermions at
$L^z$=(top) 45,80, (center) 90,93, (bottom)103,117 with $\vec{r}_0$
set to the maximum of the density, which occurs at the edge.}
\label{fermicorr}
\end{figure}

\begin{figure}[h]
\centering
\includegraphics[width=0.35\textwidth,clip]{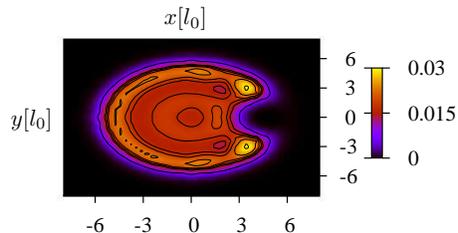}
\caption{Density-density correlation function
$\hat{\rho}(\vec{r},\,\vec{r}_0)$ of the Laughlin state for $N=12$
dipolar fermions with $\vec{r}_0$ chosen at the maximum of the
density, which occurs at the edge.}
\label{laughlin}
\end{figure}

Our results may be summarised as follows: we have studied in detail
ground and excited states of quasi-2D ultra-cold rotating dipolar
Fermi gases.  By exact diagonalization methods, we studied systems up
to 12 particles.  We have identified novel kinds of strongly
correlated states in the intermediate regime, i.~e., "boosted"
pseudo-hole ground states, which appear alternatively as $\Omega$
grows (see Fig.~\ref{fermicorr}). The calculation of the substantial
gap in the excitation spectrum of the dipolar Laughlin state at
$\nu=1/3$ (see Fig.~\ref{laughlin}) proves the accessibility of
fractional quantum Hall states in these microsystems.  At lower
fillings, interactions favour crystalline order.  Rotating dipolar
gases are thus very suitable candidates to realize Laughlin-like and
more exotic quantum liquids, as well as their crossover behaviour to
Wigner crystals.

\paragraph{\bf Wigner crystals}

\begin{figure}[h]
\begin{center}
\includegraphics[width=7.25cm]{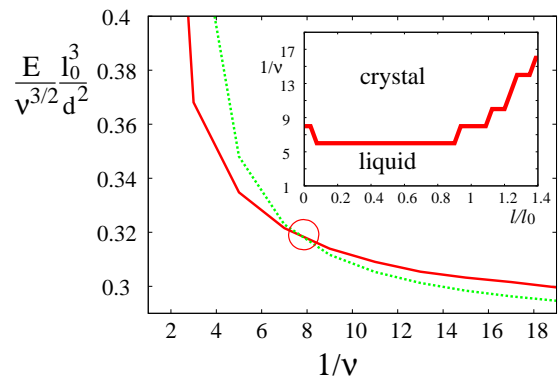} 
\end{center}
\caption{Energy per particle for the Wigner crystal (dotted line) and for
the Laughlin state (solid line) as a function of the filling factor for $l=0$%
. Insert shows the critical filling factor as a function of the extension in
the $z$-direction.}
\label{transition}
\end{figure}

Finally, in Ref.~\cite{Fehrmann:2006aa} we have discussed the
existence of a Wigner crystal phase in a rapidly rotating gas of
polarised dipolar fermions. It is shown that for sufficiently low
filling factors $\nu <1/7$ the Wigner crystal has a lower energy than
the Laughlin liquid (see Fig.~\ref{transition}). We have also examined
the stability of the Wigner crystal state by incorporating
phonon-phonon interactions and identified the quantum melting point with
the appearance of imaginary frequencies. For sufficiently high
magnetic fields the critical filling factor is a constant and we
formulate the Lindemann criterion. Note that in the considered case of
a rotating dipolar gas, the crystal phase exists at lower densities,
contrary to a non-rotating gas, in which the ground state has a
crystal order at high densities, see Ref.  \cite{Lozovik}. In fact
detailed studies of this latter possibility using quantum Monte Carlo
methods have been performed by the group of J. Boronat in Barcelona,
together with G. Astrakharchik.  These authors have in particular
studied weakly interacting two-dimensional systems of dipoles and
limitations of mean-field theory \cite{grig1}, Wigner cristallization
and quantum phase transition in a two-dimensional system of dipoles
\cite{grig2}, and also 1D dipolar gases \cite{grig3}.

\vfill

\clearpage

\paragraph{\bf Conclusions}

We conclude with the standard conclusion of M. Lewenstein and the
motto to the review \cite{ml}, which is a citation from William
Shakespeare's "Hamlet":

\begin{center}
\noindent{\it There are more thing in heaven and earth, Horatio,
than are dreamt of in your philosophy.}
\end{center}

\noindent In the present context, it expresses our never ending
curiosity, enthusiasm, joy, and excitement of working in atomic
physics in general, and physics of ultra-cold dipolar atoms in particular.

We thank all the co-authors of the papers on dipolar gases and related
subjects. We acknowledge support from EU IP Programme "SCALA", ESF
PESC QUDEDIS, MEC (Spanish Government) under contracts FIS 2005-04627,
Consolider Ingeni 2010 ``QOIT'', and Acciones Integradas
ICFO-Hannover. C.M. acknowledges financial support by the EU through
an EIF Marie-Curie Action.

\end{document}